\documentclass[secnumarabic,amssymb,  aps, prl, reprint, nobibnotes]{revtex4-1}

\usepackage{amsmath}
\usepackage{graphicx}
\usepackage{bm}
\usepackage[usenames,dvipsnames]{color}
\usepackage{epstopdf}

\begin{document}

\title{Translocation frequency of double-stranded DNA through a solid-state nanopore}

\author{Nicholas A. W. Bell$^{1}$}
\author{Murugappan Muthukumar$^{1,2}$}
\author{Ulrich F. Keyser$^{1}$}
\affiliation{$^{1}$Cavendish Laboratory, University of Cambridge, CB3 0HE, UK}
\affiliation{$^{2}$Polymer Science and Engineering Department, University of Massachusetts, Amherst, Massachusetts 01003, USA}

\begin{abstract}
Solid-state nanopores are single molecule sensors that measure changes in ionic current as charged polymers such as DNA pass through.  Here, we present comprehensive experiments on the length, voltage and salt dependence of the frequency of double-stranded DNA translocations through conical quartz nanopores with mean opening diameter 15 nm. We observe an entropic barrier limited, length dependent translocation frequency at 4M LiCl salt concentration and a drift-dominated, length independent translocation frequency at 1M KCl salt concentration. These observations are described by a unifying convection-diffusion equation which includes the contribution of an entropic barrier for polymer entry. 
\end{abstract}

\maketitle

\begin{center}
 \textbf{I. INTRODUCTION}
\end{center}
Polymer translocation through a narrow pore is a ubiquitous process in organisms.  Classic examples include the passage of mRNA across the nuclear pore complex and DNA ejection by a bacteriophage.  $\emph{In vitro}$ experiments on polymer translocation are possible using nanopores fabricated in solid-state materials or with reconstituted membrane proteins.  The basic method for sensing the translocation of a polymer, such as DNA, relies on applying a voltage across the nanopore and measuring changes in ionic current as molecules pass through. This simple premise underlies research for developing nanopores as biosensors for instance in next-generation sequencing applications \cite{Branton2008}.  Yet experimental data on polymer translocation phenomena present many puzzles that remain to be fully understood \cite{Muthukumar2009}.

Conceptually, the electrophoretically driven translocation of DNA through a nanopore consists of essentially three steps; (i) the DNA moves towards the pore entrance by a combination of diffusion and drift due to the electric field outside the pore, (ii) the DNA is captured at the mouth of the pore, and (iii) the DNA threads through the nanopore causing a detectable ionic current change. All three steps taken together determine the translocation frequency defined here as the number of DNA strands passing through the nanopore per second per unit concentration.  Accurate measurements of the translocation frequency enable characterisation of the transport process and are important for applications in nucleic acid sensing.

For the $\alpha$-hemolysin nanopore, the translocation frequency of short, single-stranded (ss)DNA shows an exponential dependence on voltage under typical experimental conditions indicating a barrier-limited process \cite{Henrickson2000,Meller2002,Ambjornsson2002}. At high voltages and polymer concentrations, a second exponential is observed which is attributed to the effect of polymer-polymer interactions at the pore mouth \cite{Meller2002}. Small diameter (sub 5-nm) solid-state nanopores show an increasing translocation frequency with double-stranded (ds)DNA length from 0.4 kbp to $\sim$ 8 kbp followed by an indication of a length independent translocation frequency for longer lengths \cite{Wanunu2010b}.  Electro-osmotic flow effects on the dsDNA trapped at the membrane surface were suggested as a reason for the length dependence of translocation frequency \cite{Wanunu2010b,Grosberg2010}.  It is known that the small diameter of these nanopores relative to the dsDNA cross-section results in strong interactions between the dsDNA and pore surface \cite{Wanunu2008,VandenHout2010a}. Larger nanopores, with diameters several times the cross section of dsDNA, show significantly less spread in transit times due to the lack of strong surface interactions \cite{Wanunu2008}. This enables a simpler physical picture with the translocation speed accurately modelled by a combination of hydrodynamics and continuum electrostatics \cite{Ghosal2006,Ghosal2007a}.  While many reports in the literature have characterised the threading speed and conformations of dsDNA in these wide nanopores \cite{Mihovilovic2013,Storm2005,Li2003} there are few measurements on the translocation frequency which can provide insight on the underlying transport mechanisms.  

In this paper, towards developing a complete picture of dsDNA translocation in solid-state nanopores, we study the voltage, length, salt and concentration dependence of dsDNA translocation frequency for nanopores with diameters 15$\pm$3 nm.  We show the simultaneous detection of ten different DNA lengths in the range 0.5 kbp to 10 kbp which allows us to accurately characterize the translocation frequency.  In 4M LiCl electrolyte, there is an increase in translocation frequency with DNA length whereas experiments using 1M KCl show a length independent translocation frequency.  These trends are accurately captured by a 1D convection-diffusion equation which incorporates an entropic barrier for polymer entry into the nanopore and scaling laws for polymer diffusion and mobility.  
\begin{center}
 \textbf{II. METHODS}
\end{center}
Quartz capillaries were fabricated into nanopores using a previously published protocol with diameters 15 $\pm$ 3 nm (mean$\pm$s.d.) \cite{Bell2015}.  The nanopore was filled with a solution of 10 mM Tris-HCl pH 8, 1 mM EDTA and electrolyte (either 4M LiCl or 1M KCl). DNA was added to the reservoir outside the nanopore tip which was set as the electrical ground.  Table I gives the final reservoir concentrations of each DNA strand used for the experiments in 4M LiCl in Figures 2 and 4. The sample was made by diluting a mix of DNA lengths (1 kb DNA ladder) purchased from New England Biolabs.  For experiments using 1M KCl, three DNA lengths of 5 kbp, 10 kbp and 20 kbp (NoLimits chromatography-purified DNA, ThermoFisher Scientific) were mixed in an equimolar ratio
so that the final reservoir concentration was 0.12 nM of each length.

Fig.~1a shows a typical scanning electron microscope image of a conical quartz nanopore and in Fig.~1b the electric field profile at +600 mV applied potential is shown from finite-element calculations.  The electric field was calculated using a finite-element solver (COMSOL 4.4) and the full Poisson-Nernst-Planck equations for ion transport through the pore in a 2D axisymmetric geometry.  The electric field is mainly confined to the last few hundred nanometres of the tip and decays rapidly away from the pore entrance.  Ionic currents were measured using an Axopatch 200B amplifier, filtered at 50~kHz using an external 8-pole Bessel filter and subsequently sampled at 250 kHz.  All data was analysed with a custom-written program in Labview 2013 (National Instruments). All experiments were performed at 20$^{\circ}$C. Translocations were detected using a threshold analysis with minimum current change set to 50~pA, minimum translocation duration 50~$\mu$s and minimum event charge deficit of 3~fC.   The translocations from all DNA lengths at all voltages fall outside these limits and therefore we do not miss translocations (Figure 15).

\begin{table}
\begin{tabular}{lllllllllll}
\hline
\multicolumn{1}{|l|}{\begin{tabular}[c]{@{}l@{}}DNA \\ length (kbp)\end{tabular}} & \multicolumn{1}{c|}{0.5} & \multicolumn{1}{c|}{1}   & \multicolumn{1}{c|}{1.5}  & \multicolumn{1}{c|}{2}    & \multicolumn{1}{c|}{3}    & \multicolumn{1}{c|}{4}    & \multicolumn{1}{c|}{5}    & \multicolumn{1}{c|}{6}    & \multicolumn{1}{c|}{8}   & \multicolumn{1}{c|}{10}   \\ \hline
\multicolumn{1}{|l|}{Conc (nM)}                                                   & \multicolumn{1}{c|}{3.2} & \multicolumn{1}{c|}{1.6} & \multicolumn{1}{c|}{0.91} & \multicolumn{1}{c|}{0.91} & \multicolumn{1}{c|}{1.6} & \multicolumn{1}{c|}{0.31} & \multicolumn{1}{c|}{0.32} & \multicolumn{1}{c|}{0.32} & \multicolumn{1}{c|}{0.20} & \multicolumn{1}{c|}{0.16} \\ \hline

\end{tabular}
\caption{Final reservoir concentration of DNA lengths in ladder used for Figures 2 and 4.} 
\end{table}

\begin{figure}
 \includegraphics{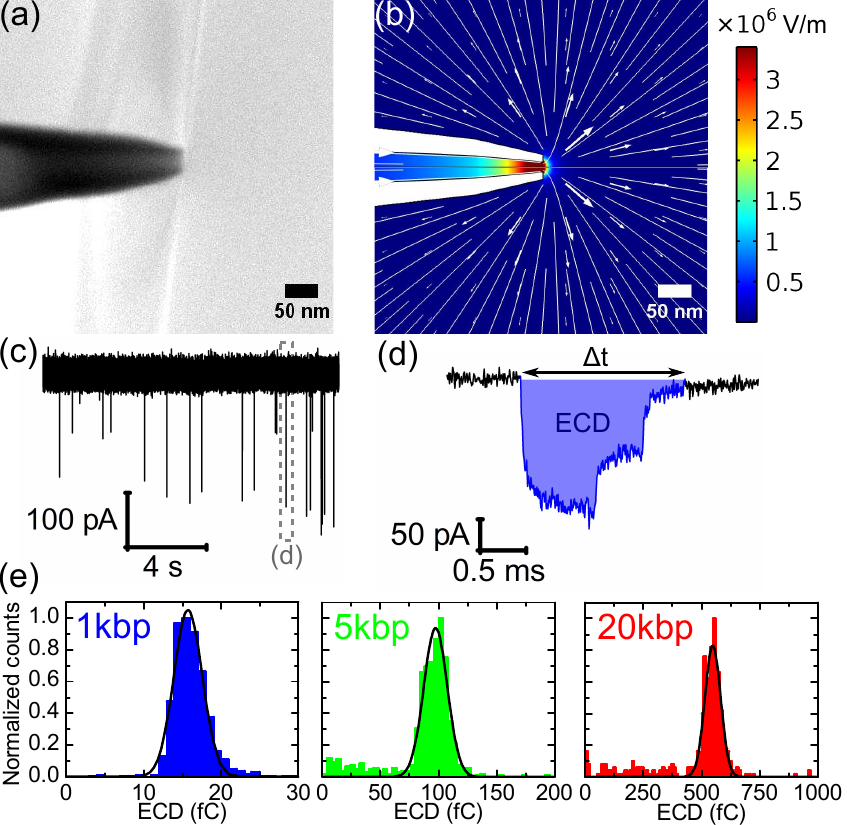}
 \caption{\label{Figure1} (a) Scanning electron microscope image of a typical quartz nanopore.  (b) Finite-element analysis of the electric field profile at a voltage of +600 mV using the geometric model given in  \cite{Bell2015}.  Lines represent tangents to the electric field.   (c) Typical current trace showing downward spikes due to DNA translocation interrupting the baseline. The event in the dashed area is shown at greater time resolution in (d) which also shows the event charge deficit (ECD) in blue - the integrated area of the event relative to the baseline. (e) Histograms of ECD for 1 kbp, 5 kbp and 20 kbp DNA with N=1119, N=764 and N=456 translocations respectively. Each length was measured separately at +600 mV with a DNA concentration of 1.6 nM.}
 \end{figure}

\begin{center}
 \textbf{III. RESULTS}
\end{center}
Initially we characterised the translocation of individual dsDNA lengths in 4M LiCl electrolyte. We observe that dsDNA can translocate in folded configurations in agreement with previous studies using similar sized nanopores \cite{Storm2005,Steinbock2010a}. This is shown by the quantised levels of current blockade as for example in Fig.~1d. For 10 kbp DNA we observe folded configurations with up to three double-strands of the same molecule in the pore (Figure 7).  The translocation time is determined by both the DNA length and the folding configuration of the DNA and does not uniquely identify a \emph{single} DNA length.  However, the integral of the current with respect to time or event charge deficit (ECD), can be used to distinguish each DNA length.   Fig.~1e shows ECD distributions of individual dsDNA lengths of 1 kbp, 5 kbp and 20 kbp which are single peaks each well fitted by a Gaussian function (Fig.~1e).   Outliers are observed at low ECD values for both 5 kbp and 20 kbp which we attribute to  a small amount of fragmentation from DNA shearing which can occur during pipetting of long DNA molecules \cite{Lengsfeld2002} and has been observed previously \cite{Mihovilovic2013}.  

Our observation of a single peak in ECD for each DNA length strongly suggests that we only measure a current signal above the background noise when the molecule fully translocates through the nanopore. It is also possible that there are events where the DNA is captured at the pore mouth but does not overcome the energy barrier to translocation and therefore returns to the sample reservoir under thermal motion.  These collisions can create a second population and have been observed for dsDNA in sub-5~nm solid-state nanopores \cite{VandenHout2010a, Wanunu2008}, in a pressure-voltage trap system for 10~nm diameter nanopores \cite{Hoogerheide2014} and for ssDNA in $\alpha$-hemolysin \cite{Kasianowicz1996}.  We attribute the lack of a second population from collisions as being due to the conical shape of the nanopores meaning that the effective sensing length is on the order of a few hundred nanometres.  Therefore collisions do not cause a significant enough current change to be measured above the noise.

\begin{figure}
 \includegraphics{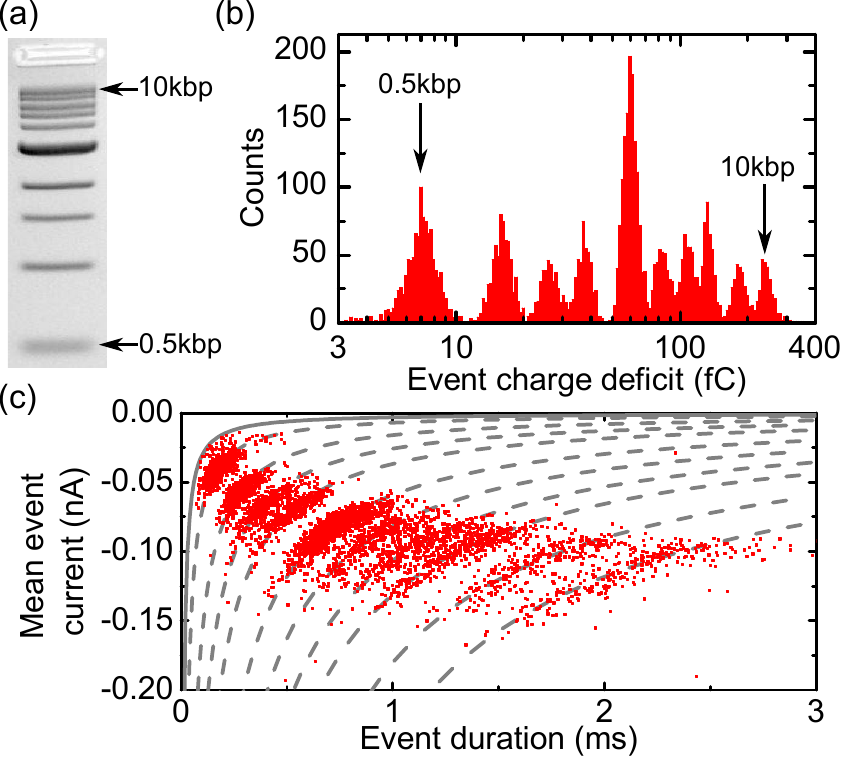}
 \caption{\label{Figure2} (a) Agarose gel (1$\%$) electrophoretogram of the DNA sample containing 10 lengths in the range 0.5 kbp to 10 kbp.   (c)  ECD histogram for 5711 translocation events recorded at +600 mV.   The histogram is logarithmically binned and the peaks corresponding to the shortest and longest DNA lengths are labelled. (c)   Scatter plot of mean event current against event duration for same events as in (b). The dashed grey lines are fits according to $\Delta I_{Mean}= -\frac{ECD_{Peak}}{\Delta t}$ and the solid line shows the 3 fC threshold.  }
 \end{figure}

For accurate, simultaneous determination of translocation frequency of a range of lengths, we used a mixture of 10 DNA lengths in the range 0.5 kbp to 10 kbp (Fig.~2a). The ECD histogram  for 5711 translocations with 4M LiCl electrolyte is shown in Fig.~2b and Fig.~2c shows the corresponding scatter plot of mean current versus duration for each translocation.  The low noise of quartz nanopores combined with the slow DNA velocity in 4M LiCl means that we have sufficient resolution to identify the 10 DNA lengths present with each DNA length in the scatter plot observed as a band. The dashed grey lines are given by  
$\Delta I_{Mean}= -\frac{ECD_{Peak}}{\Delta t}$ where $\Delta I_{Mean}$ is the mean event current, $\Delta t$ is the event duration and $ECD_{Peak}$ is the centre of the Gaussian function determined by fitting each ECD peak.  The excellent fit of this equation to the data confirms that for all translocations of a particular length, the ECD value is constant irrespective of the folding state of the DNA as it passed through. As described earlier, we assign each data point to a translocation of DNA and therefore determined the total number of translocations for each length from the integrated area of Gaussian functions fitted to each peak in the ECD histogram. The translocation frequency (shown in Fig.~4) was then calculated as $N/c\tau$ where $N$ is the number of translocations, $c$ is the DNA concentration and $\tau$ is the experiment time.  

\begin{center}
 \textbf{IV. TRANSLOCATION FREQUENCY MODEL}
\end{center}
We use the following model to describe the DNA translocation frequency based on Muthukumar \cite{Muthukumar2010}. There are three basic steps for the translocation process. In the first step, the DNA molecule undergoes drift and diffusion as a whole molecule with its allowed conformational fluctuations towards the mouth of the nanopore where it gets caught. At the end of this capture stage, the effective diameter (twice the radius of gyration) of the DNA molecule  is much bigger than the 15 nm mean diameter of the nanopore. Therefore, in the next step of successful translocation, the DNA chain can put only a few segments and not the whole chain. This step (second step) is entropically unfavorable due to the restriction of the number of conformations of a translocating DNA molecule. Hence the DNA molecule suffers from a free energy barrier. Also, since the pore diameter is larger than the thickness of DNA (2 nm), but not much bigger, several DNA segments can penetrate the mouth of the pore simultaneously. Depending on the particular manner in which these few segments are placed at the pore mouth, the values of the corresponding entropic barriers can vary \cite{Muthukumar2009, Wong2008}. Since several confined conformations of DNA are allowed, unlike in single-file translocation, it is reasonable to assume an effective free energy barrier for the whole chain. 

It would be of course desirable to perform simulations to compute the particular values of the free energy barriers for various particular configurations of tails versus loops and multiple loops of DNA at the pore mouth. Since this is much beyond the scope of the present status of the subject, we approximate the net effect in an ensemble of translocation events as an effective barrier. Once this second step of nucleation is successful, the third step of threading away the molecule is down-hill in free energy. In view of these considerations, we model a stochastic one-dimensional process of the polymer negotiating a free energy landscape where the first step is governed by diffusion and electrophoretic drift in bulk, the second step by an effective entropic barrier per chain, and the third step by the driving electric field. The net flux of molecules is determined by the entropic barrier, diffusion of the DNA and the electrophoretic force acting on the DNA. We can neglect electro-osmotic flows since measurements using optical tweezers show that the flow rate decreases strongly with salt concentration and is insignificant  compared to the electrophoretic force at the high salt concentrations used here \cite{Laohakunakorn2013,Laohakunakorn2015}.

The flux, $J(x,t)$, is therefore given by a 1D convection-diffusion equation:

\begin{equation}
J(x,t) = -D \frac{\partial c(x,t)}{\partial x} - c(x,t) \mu \frac{\partial V(x)}{\partial x} - \frac{Dc(x,t)}{kT} \frac{\partial F(x)}{\partial x},
\end{equation}

\noindent where $D$ is the diffusion coefficient, $x$ is the centre of mass coordinate of the DNA, $c(x,t)$ is the concentration, $\mu$ is the electrophoretic mobility, $V(x)$ is the electric potential, $T$ is the temperature and $F(x)$ is the entropic barrier.  We model the geometry as cylindrical with effective length $L$ which includes a small distance in the donor compartment where the electric field is strong enough to trap the DNA (Fig.~1b). The entropic barrier is triangular with height $u_0$ at a distance $\eta L/2$ (Fig.~3).  $\eta$ therefore describes the distance the entropic barrier extends.  We estimated $\eta$=0.25 based on assuming that the entropic barrier extends 50~nm (approximately one persistence length of dsDNA) and assuming an effective length of $L=200~nm$.  The concentration is $c_0$ at the entrance and zero at the exit.  

For single file DNA threading through a nanopore, theoretical calculations suggest a weak $N^{-0.2}$ dependence of barrier height $u_0$ on DNA length $N$  \cite{Kumar2009}. However, the nanopore diameter used here allows for non-single file threading for which the length dependence of the entropic barrier has not been calculated and we assume a barrier height independent of the DNA length. Naturally, a more complicated profile for the nanopore geometry and entropic barrier can be used but we choose this one since our goal is to capture the essential physics of the transport process and our chosen profile is readily solved analytically. 

\begin{figure}
 \includegraphics{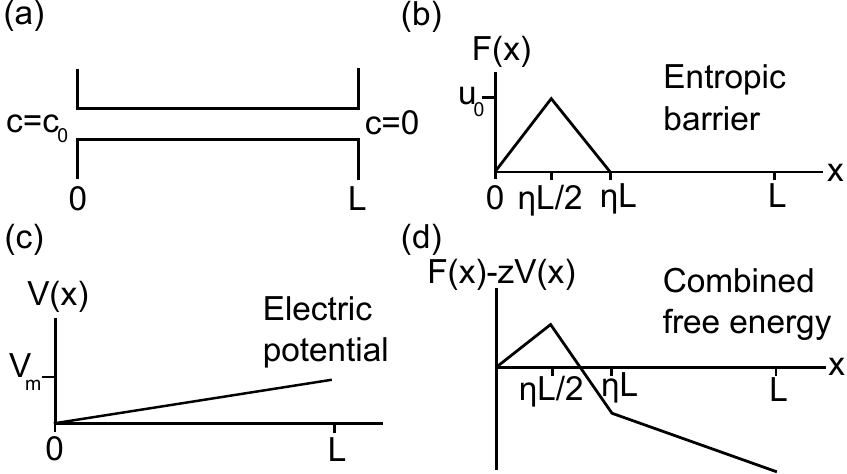}
 \caption{\label{Figure3} (a) Modelled 1D geometry with length $L$, concentration $c_0$ in the sample reservoir and zero in the opposing reservoir.  (b) Triangular profile for the entropic barrier $F(x)$ with height $u_0$.  $ \eta $ parameterises how far the barrier extends.  (c) The electric potential ($V_m$ is the applied voltage) and (d) combined free energy profile for the nanopore where $z$ is the effective DNA charge.}
 \end{figure}

By applying steady-state boundary conditions, we calculate the translocation frequency as the flux at the pore exit from equation (1). We assume the diffusion coefficient and electrophoretic mobility scale with length according to relationships observed in bulk solution.  The diffusion coefficient, $D$, of dsDNA in the length regime 0.5 kbp to 10 kbp scales with the chain length as $D \sim N^{-0.6}$ \cite{Sorlie1990,Robertson2006}. It is also known that electrophoretic mobility is independent of $N$, $\mu \sim N^0$ \cite{Stellwagen1997}. These dependencies are included in the calculations and give the length dependence for the translocation frequency (see Appendix). Our model is appropriate for dilute regimes where there are no significant polymer-polymer interactions. We performed controls at significantly diluted DNA concentrations which showed a similar dependence of DNA translocation frequency on length confirming that we are in this dilute regime (Figures 10-12).  

\begin{figure}
 \includegraphics{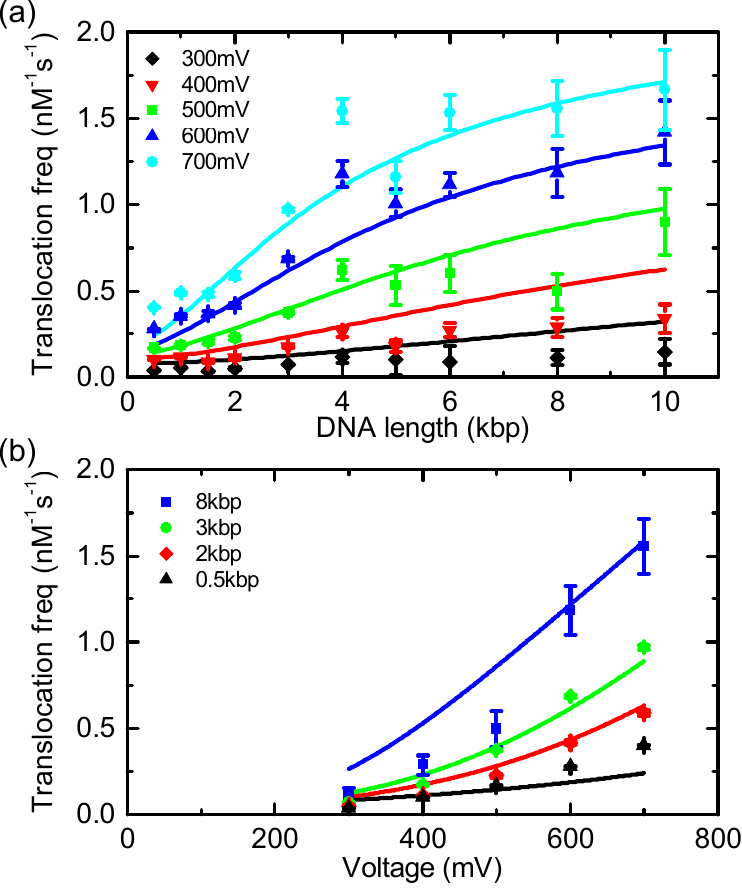}
 \caption{\label{Figure4} (a) Length dependence of DNA translocation frequency measured at five different voltages (all data from the same nanopore) in an electrolyte of 4M LiCl.  The trend lines show least-squares fits to the data using equation (11) (derived from equation (2) and the model of Fig.~3).  (b) Corresponding translocation frequency against voltage for four selected lengths.  Error bars are from the standard errors of the Gaussian fits to the ECD histogram.}
 \end{figure}

Fig.~4a shows our experimental data for the length dependence of translocation frequency in 4M LiCl.  We observe that the translocation frequency increases with increasing DNA length with an offset and steepness determined by the applied voltage.  The trend lines shown are from a least-squares fit to the data in Fig.~4a using equation (11).  Three parameters; i) the diffusion coefficient divided by effective pore length, ii) effective DNA charge and iii) entropic barrier height were globally fitted to the data set and converge close to estimates from literature with an entropic barrier height $u_0 = 4 \ kT$ (see Appendix).  Comparison of the data and fit in Fig.~4a shows that the fit deviates for long lengths at low voltage and for short lengths at high voltage.  However, the model accurately captures the trend of increasing translocation frequency with length and the changing gradient with voltage.  

Fig.~4b shows the same data as Fig.~4a plotted as translocation frequency against voltage for four selected lengths.  For 0.5 kbp, 2 kbp and 3 kbp the translocation frequency is an increasing non-linear function in voltage and for 8 kbp it is unclear whether the translocation frequency increases linearly or slightly non-linearly over the voltage range.  The model fits show a non-linear increase for 0.5 kbp, 2 kbp and 3 kbp in agreement with the data and for 8 kbp show a transition from non-linear to linear though this transition is not clearly observable in the data due to the larger error bars for 8 kbp.  

In general the model predicts two regimes of (1) entropic barrier dominated and (2) drift dominated transport \cite{Muthukumar2010}.  The entropic barrier regime dominates for low voltages and small dsDNA lengths and is characterised by a non-linear increasing translocation frequency with voltage and increasing translocation frequency with length.  The drift regime occurs at high voltage and for long dsDNA lengths where equation (1) reduces to give a translocation frequency 
\begin{equation}
\frac{J(L)}{c_0} = \frac{ \mu V_m}{L},
\end{equation}
\noindent where $V_m$ is the applied voltage and therefore the translocation frequency is linear in voltage and independent of dsDNA length.  The results presented here in 4M LiCl show the characteristics of the entropic barrier regime.  At higher voltages in Fig.~4a it can be seen that the translocation frequency is only slightly increasing with length for the longer molecules which suggests that the longer DNA lengths at higher voltage are on the threshold of the drift-dominated regime.  

The transition point between barrier-limited and drift-dominated regimes is expected to be sensitive to a number of experimental factors namely pore size and geometry, DNA effective charge (via the type of salt and the salt concentration) and temperature. The  4M LiCl electrolyte reduces DNA velocity 10 times compared to 1M KCl (a commonly used electrolyte for nanopores) which can be related to a 10 fold decrease in the effective DNA charge \cite{Kowalczyk2012a}.    We also performed experiments using 1M KCl as the electrolyte where the experimental resolution is limited to 5 kbp due to the faster translocation speed (Fig.~5a).  A linear voltage dependence of translocation frequency is observed for 5 kbp, 10 kbp and 20 kbp DNA with no significant difference between the translocation frequency for these three lengths indicating a drift-dominated behaviour in this length regime and salt concentration (Fig.~5b).  The lack of a significant barrier for translocation can also be observed by extrapolating the straight line fit to the data which shows that it intercepts close to the origin.  Our results therefore show that the threshold voltage for drift-dominated transport is significantly decreased in 1M KCl compared to 4M LiCl due to the higher effective charge.  These results at 1M KCl are supported by experiments with similar diameter nanopores which showed a linear dependence for $\lambda$-DNA (48.5 kbp) translocation frequency on voltage at 1M KCl  \cite{Chen2004}. Also results for a pressure-voltage trap system, with comparable conditions of 1.6M KCl and a 10 nm diameter pore, show that there are expected to be very few unsuccessful translocation attempts for DNA lengths of a few kilobase-pair under voltage-only experiments and hence that the electrophoretic drift dominates at this regime \cite{Hoogerheide2014}.

\begin{figure}
 \includegraphics{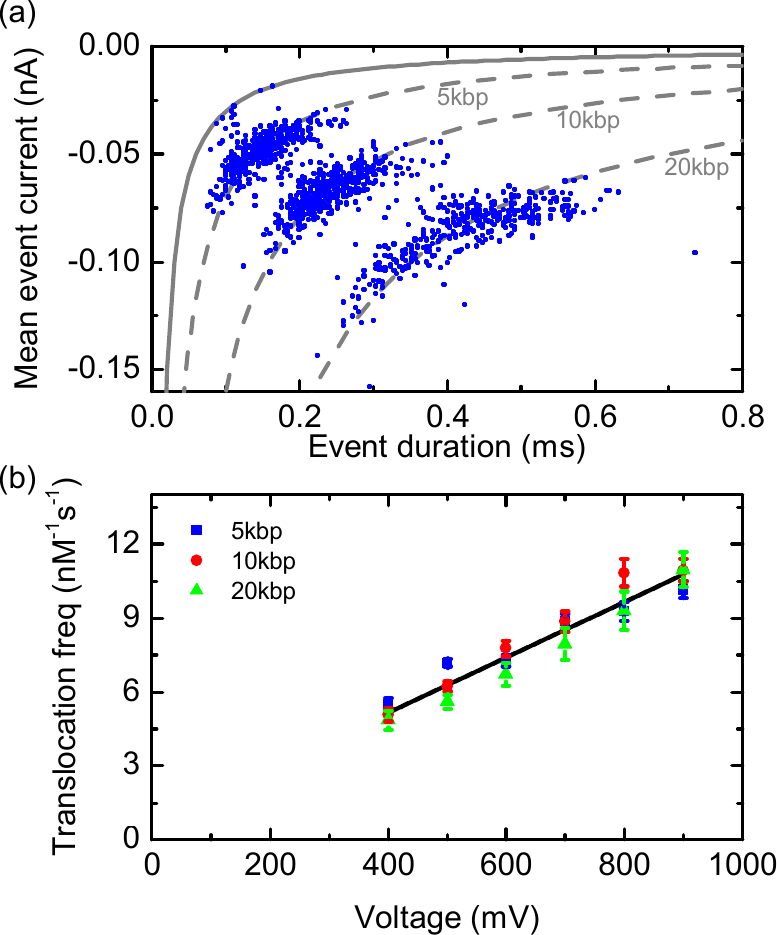}
 \caption{\label{Figure5} (a) Scatter plot of 1168 translocations from a sample containing 5 kbp, 10 kbp and 20 kbp with each length at a concentration of 0.12 nM and using 1M KCl electrolyte with +600 mV applied voltage. The dashed grey lines are fits according to $\Delta I_{Mean}= -\frac{ECD_{Peak}}{\Delta t}$ and the solid line shows the 3 fC threshold.  (b) Corresponding translocation frequency against voltage for 5 kbp, 10 kbp and 20 kbp - the translocation frequency values are significantly higher than that observed in 4M LiCl (Fig.~4b) due to the higher effective DNA charge in 1M KCl. The line shows a linear fit to all data points yielding a gradient of 0.011 $nM^{-1}s^{-1}mV^{-1}$. Error bars are from the standard errors of the Gaussian fits to the ECD histogram.}
 \end{figure}

\begin{center}
 \textbf{V. CONCLUSIONS}
\end{center}
To conclude, we have investigated the translocation frequency of dsDNA through quartz nanopores with mean diameter 15 nm.  The low noise of these nanopores, combined with the narrow distribution of dwell times for individual lengths in such wide pores, allows us to accurately measure the translocation frequency using calibrated DNA samples. At an electrolyte concentration of 4M LiCl we have observed an increasing translocation frequency as a function of DNA length in the range 0.5 kbp to 10 kbp which fits the behaviour of entropic barrier-limited transport in a 1D convection-diffusion equation and allows us to extract an approximate value of 4 kT for the entropic barrier in this model. At a lower salt concentration of 1M KCl where the effective DNA charge is higher, we measure the characteristic transport properties of a drift-dominated regime.  Our model accounts for the behaviour in these two different regimes and provides a basis for understanding and predicting polymer transport through nanopores across a wide range of conditions. 
\begin{center}
 \textbf{ACKNOWLEDGEMENTS}
\end{center}
\begin{acknowledgments}
    We thank Nadanai Laohakunakorn for finite element analysis calculations and useful discussions.  NAWB was supported by an EPSRC doctoral prize and an ERC starting grant (Passmembrane 261101), UFK acknowledges support from an ERC starting grant (Passmembrane 261101). MM acknowledges grants from the National Institutes of Health (Grant No. R01HG002776-11) and AFOSR (Grant No. FA9550-14-1-0164) and support from a Royal Society travel grant.  All data are provided in full in the results section and appendix of this paper.
\end{acknowledgments}

\begin{center}
 \textbf{APPENDIX: ANALYTICAL MODEL DERIVATION}
\end{center}

\noindent Equation (1) gives the 1D convection-diffusion equation:
\begin{equation}
J(x,t) = -D \frac{\partial c(x,t)}{\partial x} - c(x,t) \mu \frac{\partial V(x)}{\partial x} - \frac{Dc(x,t)}{kT} \frac{\partial F(x)}{\partial x} ,
\end{equation}
the mobility $\mu$ is given by 
\begin{equation}
\mu = \frac{zD}{kT} = \widetilde{z} D,
\end{equation}
where $z$ is the effective charge and $\widetilde{z} = \frac{z}{kT}$.  Also writing $\widetilde{F}(x) = \frac{F(x)}{kT}$ gives an expression for the flux of
\begin{equation}
J(x,t) = -D\left(\frac{\partial c(x,t)}{\partial x} + c(x,t) \widetilde{z} \frac {\partial V(x)} {\partial x} + c(x,t) \frac {\partial \widetilde{F}(x)} {\partial x}\right).
\end{equation}
To calculate the translocation frequency we find the steady state solution of this equation with the boundary conditions $c(x=0) = c_{0} $, $c(x=L) = 0$.  The entropic barrier $\widetilde{F}(x)$ is assumed to be triangular as shown in Figure 3 so that:
\begin{equation}
\widetilde{F}(x) = 
\left\{\begin{matrix}
2\widetilde{u_0} \frac{x}{l},  & 0<x<\frac{l}{2} \\ 
2\widetilde{u_0} (1- \frac{x}{l}), & \frac{l}{2} < x< l\\ 
0, & l<x<L 
\end{matrix}\right.
\end{equation}

\noindent where $l=\eta L$ and $\widetilde{u_0} = \frac{u_0}{kT}$.  The electrical potential in the pore is given by 
\begin{equation}
V(x) =V_m \frac{x}{L}, \; \; \; \;0<x<L
\end{equation}

\noindent where $V_m$ is the applied potential. Therefore the total free energy divided by kT, $\widetilde{W}(x)$, is given by 
\begin{equation}
\widetilde{W}(x) = 
\left\{\begin{matrix}
2\widetilde{u_0} \frac{x}{l} - \left | \widetilde{z} V_m  \right | \frac{x}{L},  & 0<x<\frac{l}{2} \\ 
2\widetilde{u_0} (1- \frac{x}{l}) - \left | \widetilde{z} V_m  \right | \frac{x}{L}, & \frac{l}{2} < x< l\\ 
- \left | \widetilde{z} V_m  \right | \frac{x}{L}, & l<x<L 
\end{matrix}\right.
\end{equation}

\noindent In steady-state the flux is independent of time and $x$ and is calculated by integrating equation (5) which gives \cite{Muthukumar2010}
\begin{equation}
J(L)= -D\frac{c(L) e^{\widetilde{W}(L)}-c(0)e^{\widetilde{W}(0)}}{\int_{0}^{L}e^{\widetilde{W}(x)}dx},
\end{equation}

\noindent and since $c(x=0)=c_0$, $c(x=L) =0$,
\begin{equation}
J(L)= D\frac{c_0}{\int_{0}^{L}e^{\widetilde{W}(x)}dx},
\end{equation}

\noindent calculating this integral with the assumed free energy profile $\widetilde{W}(x)$ gives

\begin{eqnarray}
J(L)=&& \frac{Dc_0}{L} . \bigg(\frac{1}{(\frac{2\widetilde{u_0}}{\eta}-\left | \widetilde{z} V_m \right |)}    [  e^{(\frac{2\widetilde{u_0}}{\eta} - \left | \widetilde{z}V_m \right |) \frac{\eta}{2}} -1] \nonumber\\
&& -\frac{e^{2\widetilde{u_0}}}{(\frac{2\widetilde{u_0}}{\eta} + \left | \widetilde{z}V_m \right |)} [  e^{-(\frac{2\widetilde{u_0}}{\eta} + \left | \widetilde{z}V_m \right |) \eta} -e^{-(\frac{2\widetilde{u_0}}{\eta} + \left | \widetilde{z}V_m \right |) \frac{\eta}{2}}] \nonumber\\
&&-\frac{1}{\left | \widetilde{z}V_m \right |} [e^{- \left | \widetilde{z}V_m \right |}-e^{- \left | \widetilde{z}V_m \right | \eta}]\bigg)^{-1}.     
\end{eqnarray}

When the electrophoretic force dominates over the barrier $\eta  \left | \widetilde{z} V_m \right | \gg \widetilde{u_0}$ and equation (11) reduces to 
\begin{equation} 
J(L) =  \frac{ Dc_{0} \widetilde{z} V_m}{L} = \frac{ c_{0} \mu V_m}{L},
\end{equation}
which is the same as equation (2) and is the drift-dominated regime found in 1M KCl in Figure 5.  When the drift does not dominate, we must use the full equation (11). The length dependence enters from the dependence on $D$ and $\widetilde{z}$, and we consider these in turn.
The diffusion coefficient scales according to \cite{Sorlie1990, Robertson2006},
\begin{equation}
D=D_0 N^{-0.6}.
\end{equation}
Sorlie $\emph{et al.}$ used dynamic light scattering to determine the diffusion coefficient of isolated DNA molecules in 100~mM NaCl,  10~mM Tris-HCl (pH=8), 1~mM EDTA at 20$^{\circ}$C. Their data yields a value of $D_0 = 5.9 \times 10^{-10} \ m^2 s^{-1}$ (Figure 6). 

\begin{figure}
 \includegraphics{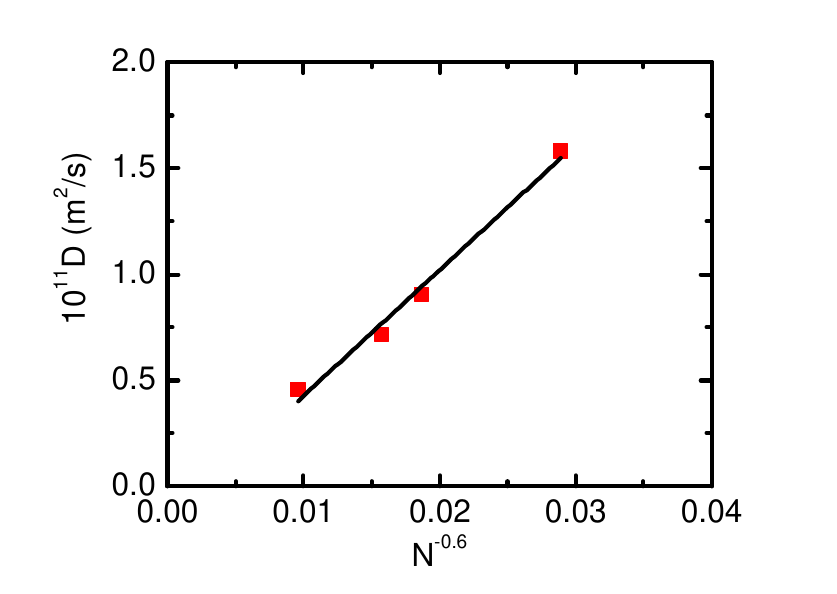}
 \caption{ Diffusion coefficient of DNA against $N^{-0.6}$ where N is in basepairs (using data from  \cite{Sorlie1990}).  A least-squares linear fit to the data is shown.}
 \end{figure}

The electrophoretic mobility of double-stranded DNA $>$ 100 bp (and other polyelectrolytes) is experimentally measured to be independent of length \cite{Hoagland1999, Stellwagen2002, Stellwagen2003, Stellwagen1997} 
\begin{equation}
\mu \sim N^{0},
\end{equation}

\noindent and therefore 
\begin{equation} 
\widetilde{z} = \widetilde{z_0} N^{0.6}.
\end{equation}

\noindent We estimate $\widetilde{z_0}$ by the following argument.  DNA has a bare charge from its phosphate groups of 2e$^-$/bp.  Experiments on tethered DNA in a nanopore at 1M KCl showed a 75\% reduction in charge comparable to the charge reduction expected from Manning condensation \cite{Keyser2006b}.  Furthermore, as reported by Kowalczyk $\emph{et al.}$ \cite{Kowalczyk2012a}, we observe a ten-fold reduction in DNA velocity for experiments in 4M LiCl compared to 1M KCl for our nanopores.  Therefore we estimate an additional factor of 10 reduction in effective charge giving a total of a 40 fold reduction from the bare charge.   Experiments are done at 20$^{\circ}$C so $kT = 4.0 \times 10^{-21} \ J$  and the total charge on 2$e^{-}$ is $3.2 \times 10^{-19} \ C$.  Therefore we obtain $\widetilde{z_0} = 2 \  CJ^{-1}$. This value of $\widetilde{z_0}$ together with that estimated for $D_0$ yields a mobility of $\mu = \widetilde{z_0} D_0 = 1.2 \times 10 ^{-9} \ m^{2} s^{-1} V^{-1}$ which is comparable to a value of $\mu = 1.8 \times 10 ^{-8} \ m^{2} s^{-1} V^{-1}$ determined for 10 kbp DNA in 1M KCl \cite{Langecker2011} given the anticipated ten fold reduction due to 4M LiCl.
 
The other parameter in our model is the barrier height.  Previously the barrier height was estimated for single chain threading to be on the order 10-20 kT \cite{Kumar2009}. The pores here allow for folded configurations as the DNA passes through and to reflect this we use initial estimates of 3-10 kT for the barrier height. Furthermore, as described in Section IV, we assume that the barrier height is independent of the DNA length. 

We fit equation (11) to the data in 4M LiCl using three parameters: $\alpha=D_0/L$ (where $L$ is the effective length taken as $L=200~nm)$, $\widetilde{z_0}$,  and $\widetilde{u_0}$ using the initial values $\alpha=0.003 \ ms^{-1}, \widetilde{z_0} = 2 \ CJ^{-1}$ and $\widetilde{u_0} = 3-10$ in steps of 1. A least-squares global fit, using the Levenberg-Marquadt algorithm (OriginPro 8.5), to Figure 4a yields $\alpha=0.061 \ ms^{-1}, \widetilde{z_0} = 0.58 \ CJ^{-1}$ and $\widetilde{u_0} = 4.3$ which converged for all values of tested values of $\widetilde{u_0}$.  The parameter $\alpha$ changes most significantly relative to the initial guesses. We note that this change can be expected given the uncertainty in the effective pore length as well as assumptions about other parameters such as the barrier width, shape and length dependence which were estimated as described in the text. Also a larger diffusion coefficient (and therefore $\alpha$) might be expected compared to the initial guess since we calculated $D_0$ based on experiments at 100~mM NaCl by Sorlie $\emph{et al.}$ whereas the nanopore experiments presented here use 4~M LiCl electrolyte and high salt concentrations can cause an increase in the polymer diffusion coefficient \cite{Tinland1997}.   We would like to note that our model requires a length dependent diffusion constant to describe the data for the DNA translocation frequency at 4M LiCl since $\mu$ and $F(x)$ are length independent. If D were also length independent, equation (1) would then have no N dependence which is contrary to our observations.  

\begin{center}
 \textbf{APPENDIX: FURTHER EXPERIMENTAL CHARACTERISATION}
\end{center}
\subsection*{Nanopore geometry}
\noindent Nanopores in this study were fabricated with the same protocol as given in Bell \emph{et al}. \cite{Bell2015} where the pore diameters were estimated using scanning electron microscopy as 15 $\pm$ 3 nm (mean $\pm$ sd).  We can also obtain an indication for the pore size by analysing the number of folded DNA translocations.  Figure 7 shows translocation events and an all points histogram from 10 kbp translocations at +600 mV (selected by ECD from the data shown in Figure 2).  There is a peak at 0 pA due to the the baseline level and the one DNA double-strand level occurs at -122 pA.  A second peak is also visible at -217 pA due to configurations where the DNA has a fold so that two double-strands are present within the pore.  A third peak is also visible confirming that the pore diameter allows at least three double-strands and is therefore consistent with a diameter of $\sim$15 nm. 
\begin{figure}
 \includegraphics{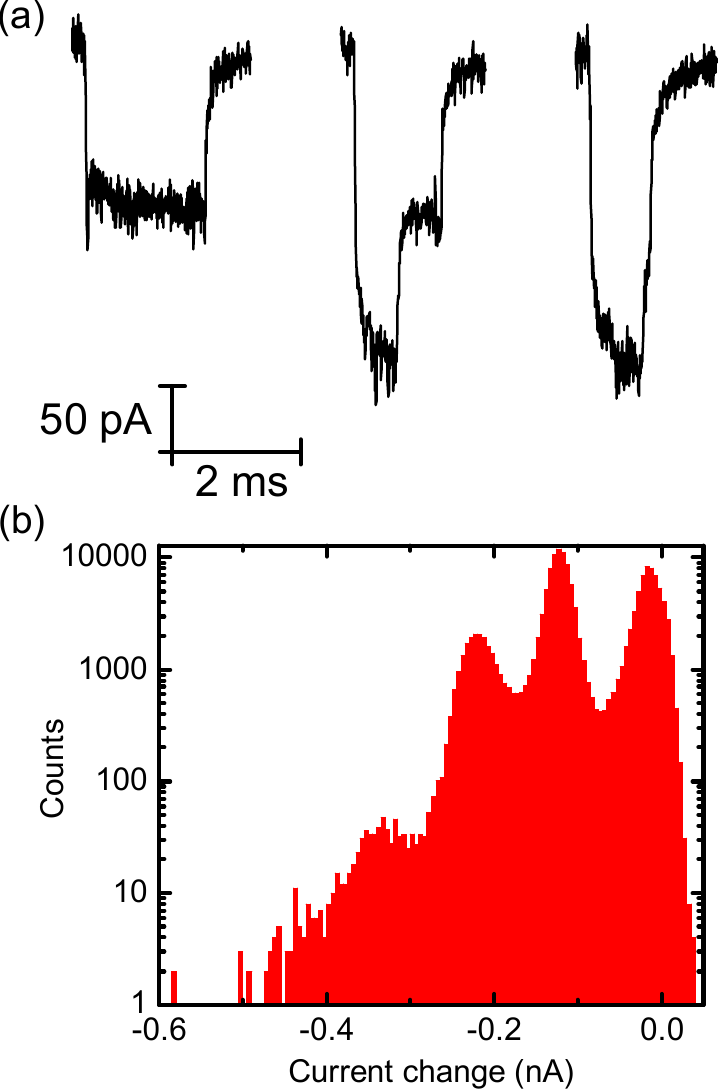}
 \caption{(a) Typical translocation events of 10 kbp DNA showing steps due to the folding configurations of the DNA as it translocates the nanopore. (b) All points histogram of 260 events showing a baseline peak at 0 nA and successive peaks due to DNA folding states.}
 \end{figure}

\subsection*{Determination of number of translocations from ECD histogram and time between successive events}

\begin{figure}
 \includegraphics{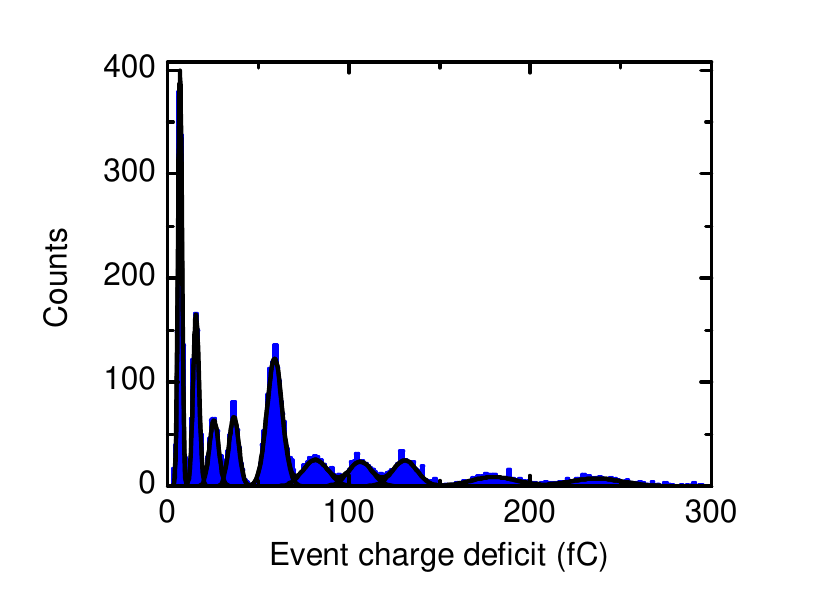}
 \caption{Example least-squares fit of multiple Gaussian functions to the 10 peaks from the 10 DNA lengths present in the sample given in Table I. The data is the same as that present in Figure 2 but with a linear rather than logarthmic binning.}
 \end{figure}

\noindent The number of DNA strands that translocated in an experiment was determined by fitting multiple Gaussian functions to a histogram of ECD values.  The fitting was done using the Levenberg-Marquadt algorithm to determine a least squares fit to the distribution (OriginPro 8.5).  The number of peaks and approximate peak positions were given as initial parameters.  Figure 8 shows an example fit to an ECD histogram data set.  The ECD range was limited between the threshold of 3 fC and $\mu$+4$\sigma$ for the final peak.  Figure 9 shows the time between successive events for the 5711 events presented in Figure 8.  We observe a clear exponential decay as expected for a Poisson process with independent DNA translocations such that the probability distribution function for a arrival in time $t$ after a translocation is 
\begin{equation}
P(t) \propto e^{-\lambda t}.
\end{equation}

\begin{figure}
 \includegraphics{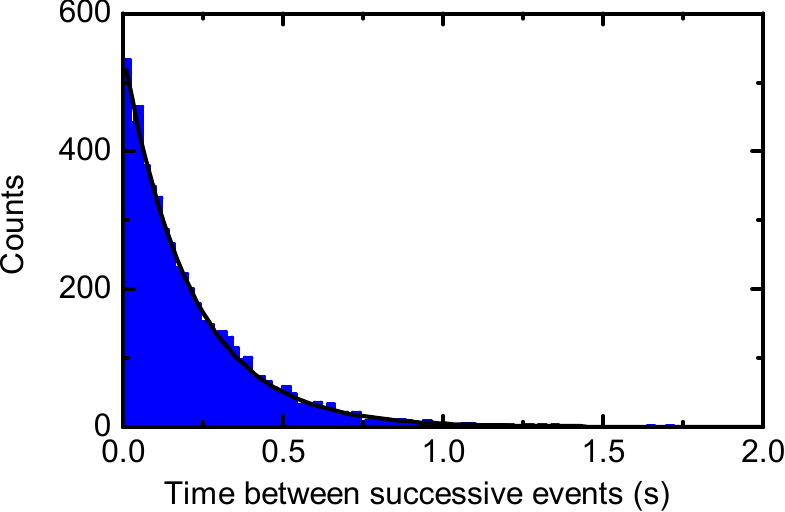}
 \caption{Histogram of time since last event for all 5711 events shown in Figure 2.  The solid line is a fit according to equation 16.}
 \end{figure} 

\subsection*{Dependence of DNA translocation frequency on concentration}
\noindent We tested the translocation frequency dependence at different concentrations relative to those given in Table I.  Figure 10 shows the translocation frequency at the concentration given in Table I (RC=1), and at three times dilution (RC=0.33).  These concentrations were tested using the same nanopore and at +600~mV applied voltage.  We observe no significant difference after dilution of the DNA sample.

\begin{figure}
 \includegraphics{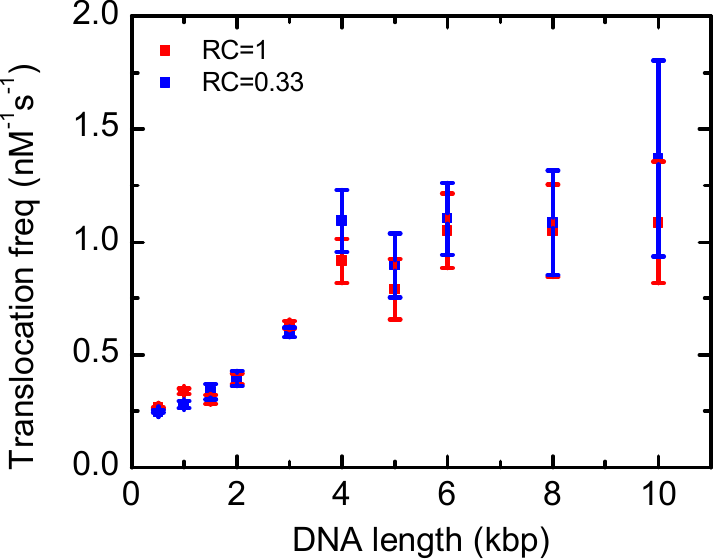}
 \caption{Dependence of DNA translocation frequency on DNA concentration. Translocation frequency against DNA length measured for an individual nanopore at the concentration given in Table I (RC=1) and at a 3 times dilution relative to Table I (RC=0.33).}
 \end{figure}

We also performed experiments on a custom-made DNA ladder to check for potential sample variability and to further extend the concentration range tested. The ladder was made by mixing 8 chromatography-purified, dsDNA lengths (NoLimits, ThermoScientific, UK).  Each DNA length concentration was measured from the absorbance at 260 nm using a Nanodrop spectrophotometer.  The DNA lengths were then mixed together and diluted in 1x TE, 4 M LiCl so that the final concentration of each DNA length in the sample reservoir was as given in Table II. Figure 11 shows an ECD histogram and scatter plot of 1066 events taken in 21 minutes of recording at +600 mV with this DNA sample. Figure 12 shows the translocation frequency of DNA corresponding to the data set in Figure 11 and an additional data set from another nanopore.  We measure the same trend as Figure 4 of an increasing translocation frequency with DNA length.  The concentration was significantly lower than that used in Table I and resulted in $\sim$1 translocation per second making it highly unlikely that there were effects due to interactions of DNA strands at the nanopore mouth. As an experimental note, we did not use lengths longer than 20 kbp in this study (such as 48.5 kbp $\lambda$-DNA) due to the difficulty in preparing accurate concentrations of such lengths.  This stems from two sources 1) an increased amount of fragments is observed with such long lengths and 2) the slow rate of mixing due to the small diffusion coefficient which means that inhomogeneous concentrations can easily arise. 

\begin{table}
\begin{tabular}{lllllllllll}
\cline{1-9}
\multicolumn{1}{|l|}{\begin{tabular}[c]{@{}l@{}}DNA \\ length (kbp)\end{tabular}} & \multicolumn{1}{c|}{1}    & \multicolumn{1}{c|}{2}    & \multicolumn{1}{c|}{3}    & \multicolumn{1}{c|}{5}    & \multicolumn{1}{c|}{7}    & \multicolumn{1}{c|}{10}   & \multicolumn{1}{c|}{15}   & \multicolumn{1}{c|}{20}   & \multicolumn{1}{c}{} & \multicolumn{1}{c}{} \\ \cline{1-9}
\multicolumn{1}{|l|}{Conc (nM)}                                                   & \multicolumn{1}{c|}{0.22} & \multicolumn{1}{c|}{0.21} & \multicolumn{1}{c|}{0.21} & \multicolumn{1}{c|}{0.19} & \multicolumn{1}{c|}{0.20} & \multicolumn{1}{c|}{0.18} & \multicolumn{1}{c|}{0.18} & \multicolumn{1}{c|}{0.18} & \multicolumn{1}{c}{} & \multicolumn{1}{c}{} \\ \cline{1-9}          
\end{tabular}
\caption{Final reservoir concentration for DNA lengths in custom DNA ladder used for Figures 11 and 12.} 
\end{table}
\begin{figure}
 \includegraphics{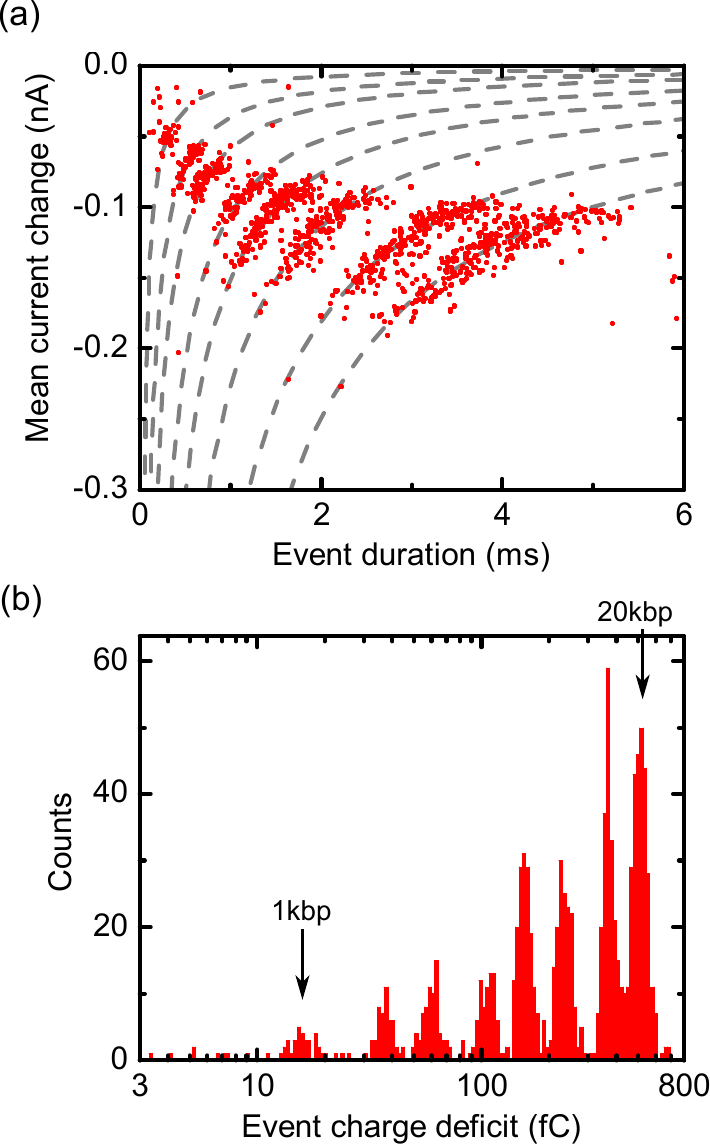}
 \caption{ Experiment on custom ladder given in Table II at 4M LiCl. (a) Scatter plot of all 1066 translocations.  Dashed lines indicate constant ECD with values determined from the centre of Gaussian peaks fitted to each ECD histogram. (b) ECD histogram of 1066 translocation events of the DNA ladder given in Table II.  Eight peaks can be distinguished corresponding to the eight DNA lengths present in the sample.}
 \end{figure}
\begin{figure}
 \includegraphics{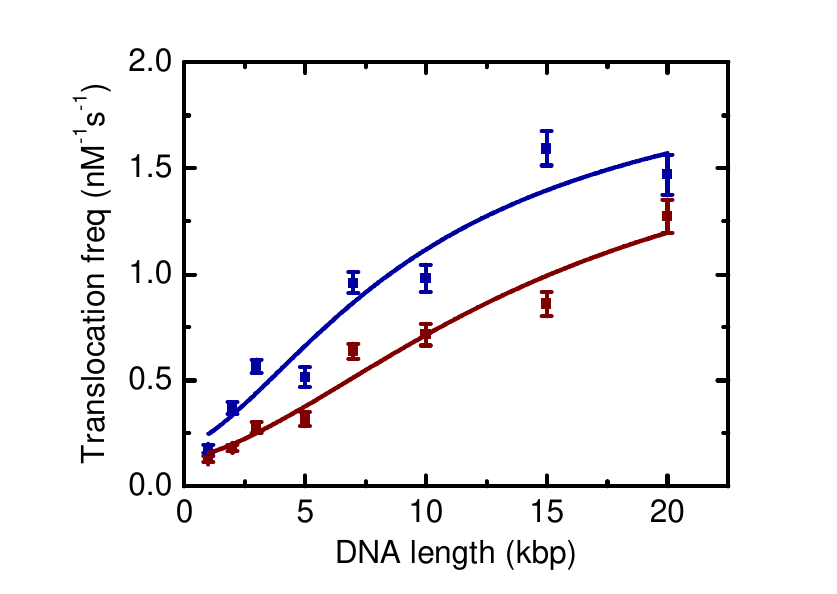}
 \caption{Translocation frequency as a function of DNA length for the DNA sample given in Table II at +600 mV.  The two colours represent data from two nanopores with fits yielding $\alpha = 0.10 \  ms^{-1}, \widetilde{z_0} = 0.38 \ CJ^{-1}$ and $\widetilde{u_0} = 4.1$ (blue data) and $\alpha = 0.12 \  ms^{-1}, \widetilde{z_0} = 0.31 \ CJ^{-1}$ and $\widetilde{u_0} = 4.6$ (brown data). Error bars represent the standard errors of the fits.}
 \end{figure}
\subsection*{Repeats of voltage dependence using different nanopores}
\noindent Figures 13 and 14 shows repeats, using different nanopores, of the voltage dependence behaviour shown in Figure 4 (using the DNA sample given in Table I).  Each nanopore can be expected to give slightly different values for translocation frequency due to small differences in geometry.  However we consistently observe, in 4M LiCl, the trend of increasing translocation frequency with increasing DNA length.  We also consistently observe that the data for 4~kbp is systematically slightly above the general trend which could be due to a slight error in the DNA concentrations or due to the fact that the 4~kbp ECD peak is next to the large 3 kbp peak originating from the high 3~kbp concentration. A slight deviation from a Gaussian peak could therefore affect the calculated 4~kbp frequency.  
\begin{figure}
 \includegraphics{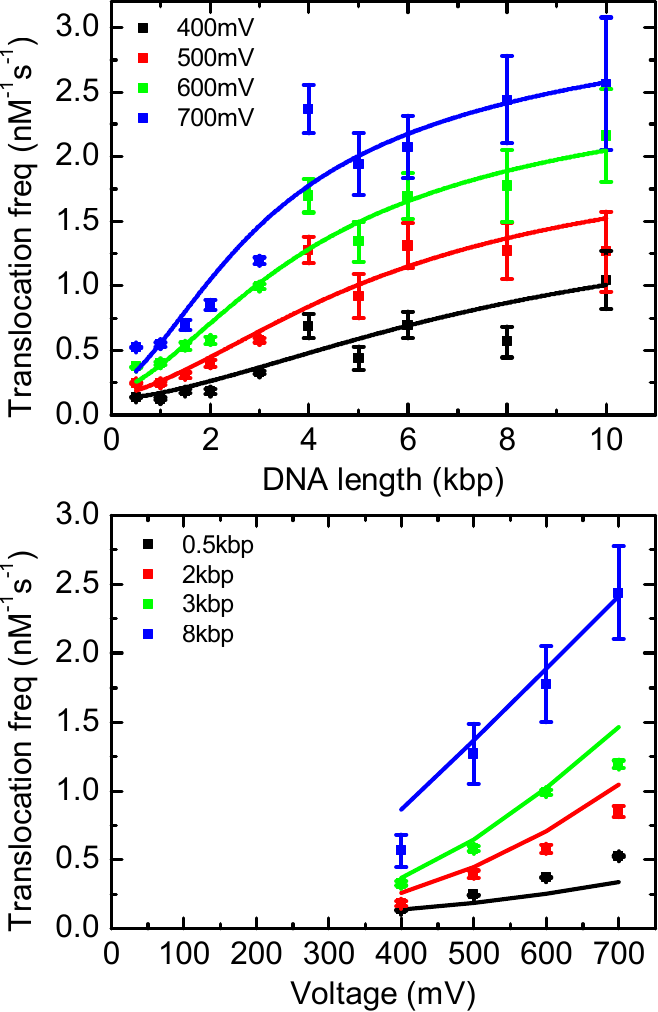}
 \caption{Repeat of Figure 4 with a different nanopore showing the consistent behaviour of increasing translocation frequency with increasing DNA length.  The global least-squares fit yields $\alpha = 0.073 \  ms^{-1}, \widetilde{z_0} = 0.68 \ CJ^{-1}$, $\widetilde{u_0} = 4.5$.}
 \end{figure}
\begin{figure}
 \includegraphics{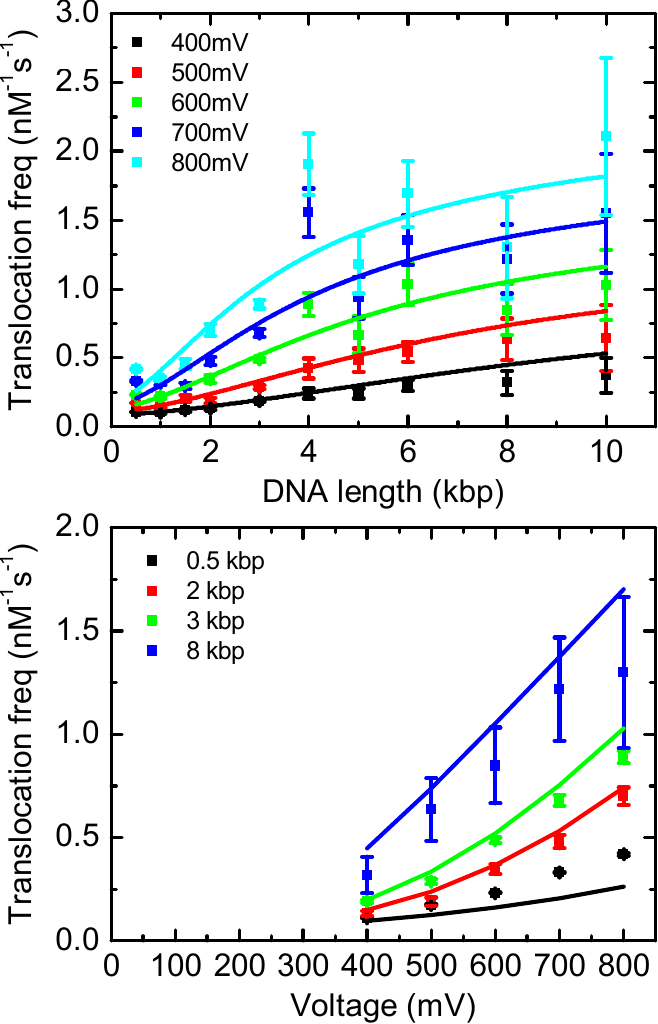}
 \caption{Additional repeat of Figure 4 with another nanopore with fits yielding $\alpha = 0.055 \  ms^{-1}, \widetilde{z_0} = 0.56 \ CJ^{-1}$, $\widetilde{u_0} = 4.3$.}
 \end{figure}

\subsection*{Translocation statistics lie oustide analysis thresholds}
\noindent The thresholds of detection of 50~pA deviation from baseline current, minimum 50~$\mu$s event duration and minimum 3~fC event charge deficit were used in all experiments. We carefully checked for each experiment that the translocations lied outside these thresholds and therefore that there are no missed translocations which would create systematic error in the calculated translocation frequency.  For example, Figure 15 shows scatter plots of mean event current and peak event current against event duration for the particular nanopore used in Figure 2 and 4.  The 50~pA and 50~$\mu$s thresholds are marked for the peak current scatter plots and the 3~fC threshold is marked for the mean event current scatter plots.  The main clusters of points for different lengths lie outside these limits therefore indicating that the DNA lengths down to 500~bp are accurately measured. 

\clearpage

\begin{figure}
 \includegraphics{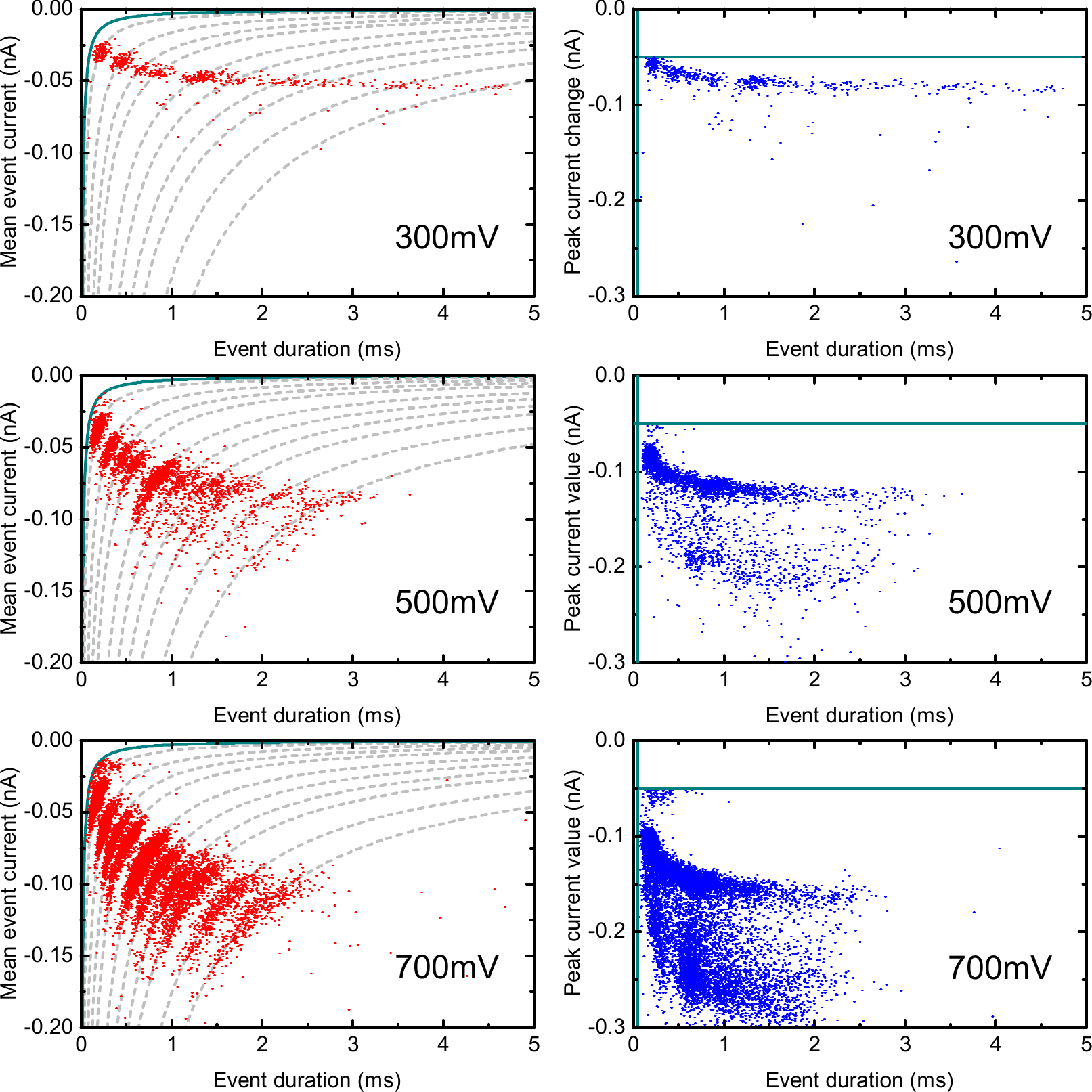}
 \caption{ Left column: Mean event current against event duration for all translocations.  The green line shows the 3~fC threshold set in the analysis.  Dashed lines indicate constant ECD with values determined from the centre of Gaussian peaks fitted to each ECD histogram.  N=637 (+300~mV), N=3174 (+500~mV) and N=11661 (+700~mV).  +300~mV and +500~mV were recorded for approximately 20 minutes and  +700~mV was recorded for approximately 30 mins.  Right column: Corresponding plots of peak current value against event duration.  The threshold limits of 50~$\mu$s and 50~pA are also marked as green lines.  Events cluster outside the threshold limits indicating that nearly no translocations are missed.}
 \end{figure}

\clearpage
\bibliographystyle{apsrev}
\bibliography{Library}

\end{document}